# *p*-orbital disclination states in non-Euclidean geometries

Ying Chen[1,#,†], Yuhang Yin[2,#], Ze-Huan Zheng[2], Yang Liu[3], Zhi-Kang Lin[3], Jing Li[2], Jian-Hua Jiang[3,†], Huanyang Chen[2,†]

[1]*College of Information Science and Engineering, Fujian Provincial Key Laboratory of Light Propagation and Transformation, Huaqiao University, Xiamen 361021, China*

[2]*Department of Physics and Pen-Tung Sah Institute of Micro-Nano Science, Xiamen University, Xiamen 361005, China*

[3]*Institute of Theoretical and Applied Physics, School of Physical Science and Technology, and Collaborative Innovation Center of Suzhou Nano Science and Technology, Soochow University, 1 Shizi Street, Suzhou, 215006, China*

[#]These authors contributed equally to this work.

[†]Correspondence should be addressed to: yingchen@hqu.edu.cn (Y.C.), jianhuajiang@suda.edu.cn (J.H.J.), or kenyon@xmu.edu.cn (H.C.).

## Abstract

Disclinations are ubiquitous lattice defects existing in almost all crystalline materials. In two-dimensional nanomaterials, disclinations lead to the warping and deformation of the hosting material, yielding non-Euclidean geometries. However, such geometries have never been investigated experimentally in the context of topological phenomena. Here, by creating the physical realization of disclinations in conical and saddle-shaped acoustic systems, we demonstrate that disclinations can lead to topologically protected bound modes in non-Euclidean surfaces. In the designed honeycomb sonic crystal for *p*-orbital acoustic waves, non-Euclidean geometry interplay with the *p*-orbital physics and the band topology, showing intriguing emergent features as confirmed by consistent experiments and simulations. Our study opens a pathway towards topological phenomena in non-Euclidean geometries that may inspire future studies on, e.g., electrons and phonons in nanomaterials with curved surfaces.

Topological defects [1–3] in crystals are the stable singularities in lattice geometry, which are usually characterized by their discrete topological charges. There are two kinds of common topological defects, the disclinations and dislocations, which are characterized by their real-space topological charges, the Frank angle and the Burgers vector, respectively [2, 4–6]. These topological defects, as real-space singularities, yield exotic effects which can be connected to artificial gauge fields and gravitational effects [7-10]. Topological defects have been used as an experimental probe of various topological phases such as topological insulators, Weyl/Dirac semimetals and higher-order topological insulators [11-27] in both two-dimensional (2D) and three-dimensional (3D) systems, to induce unique experimental features in these topological phases of matter [11, 23]. In particular, recent studies have discovered disclination-induced fractional mode charges [28, 29] and zero-dimensional (0D) [30-32] or one-dimensional (1D) [33-36] modes in photonic and acoustic topological insulators and semimetals. However, to date, all experimental studies in 2D systems focus only on planar (Euclidean) geometry [28, 32, 33], while non-planar (non-Euclidean) geometry is unexplored, although in genuine 2D materials topological defects often lead to non-Euclidean geometries [37].

Here, using a designer sonic crystal and the 3D printing technology, we create disclinations on non-Euclidean surfaces with positive or negative Gaussian curvatures which are achieved via a network of 3D acoustic cavities connected by hollow tubes in a nontrivial configuration. The designed sonic crystal yields a topological valley Hall insulator phase for acoustic waves of the $p$-wave nature due to the unique geometry of the acoustic cavities. The coexisting band topology in wavevector space and the non-Euclidean geometry due to the disclinations in real-space lead to emergent topological modes bound to the disclination core. These topological modes are observed in both acoustic pump-probe measurements and near-field wavefunction scanning. The consistent theory, simulation and experiments confirm that the interplay between non-Euclidean geometry and band topology could yield interesting phenomena that are yet to be explored in photonic, phononic and electronic systems. In particular, for electronic and phononic systems, there are plenty of non-Euclidean geometries in nanomaterials [37]. The possible nontrivial topological effects in these materials are still waiting to be investigated.

**Disclinations on non-Euclidean surfaces**

We start from a normal 2D honeycomb lattice in a flat, Euclidean plane, as shown in Fig. 1a, where the two sublattices are denoted by the yellow (A-type) and blue (B-type) spheres, respectively. The non-Euclidean disclination structures can then be constructed through a cutting and gluing procedure, which are described as follows: by removing or inserting a $2\pi/3$ sector of the perfect crystal (along the magenta dashed lines) and reconnecting the boundaries together, a conical or a saddled-shaped surface can be formed through the Kamada-Kawai layout algorithm [38], which is a force-directed graph drawing algorithms that tends to find the optimal distances between nodes in $n$-dimensional ($n \geq 2$) space. As shown in Figs. 1b and 1c, the resultant curved surface exhibits either a positive or a negative Gaussian curvature (denoted in the figure as $G > 0$ and $G < 0$, respectively) and thus forms a 3D non-Euclidean space [39, 40]. The non-Euclidean space with $G > 0$ ($G < 0$) has a quadrangle (octagon) vertex. In the following, we denote the non-Euclidean space with $G > 0$ ($G < 0$) as the conical (hyperbolic) lattice. Intuitively, these non-Euclidean geometries are related to mapping the Frank angle of the disclination from 2D to 3D spaces. We remark that in the constructed curved-space lattices, the distance between nearest-neighboring (NN) sites is kept the same as that in the flat honeycomb lattice in Fig. 1a. This property, which is enabled by the non-Euclidean geometries, is distinct from the 2D disclination structures studied in the literature which are defined in flat, Euclidean spaces [28, 32, 33].

**$p$-orbital honeycomb lattice model**

We consider spherical acoustic cavities that each supports three degenerate acoustic modes, the $p_x$, $p_y$ and $p_z$ orbitals. When many such acoustic cavities are connected by slender tubes to form a honeycomb lattice as shown in Fig. 2a, a ball-and-stick sonic crystal is realized which hosts six acoustic Bloch bands due to the $p$-orbitals. Here, the $p$-orbital physics for acoustic waves emerges due to the rich coupling geometries between the A-type and B-type sites [41]. Such couplings can be classified into the σ-type (longitudinal) couplings $t_\sigma$ and the π-type (transverse) couplings $t_\pi$ [41, 42]. Here, due to the special structure of spheres connected by tubes, the π-type couplings are much weaker and can be neglected (see Supplementary Note 1). For this reason, the $p_z$-orbitals have vanishing couplings with the $p_x$- and $p_y$-orbitals in flat (Euclidean) geometry, while such couplings remain negligible in non-Euclidean geometries. Therefore, the dispersive acoustic Bloch bands here are dominated by the σ-type couplings among the $p_x$- and $p_y$-orbitals at the neighboring sites, while the $p_z$-orbitals lead to two nearly flat acoustic bands.

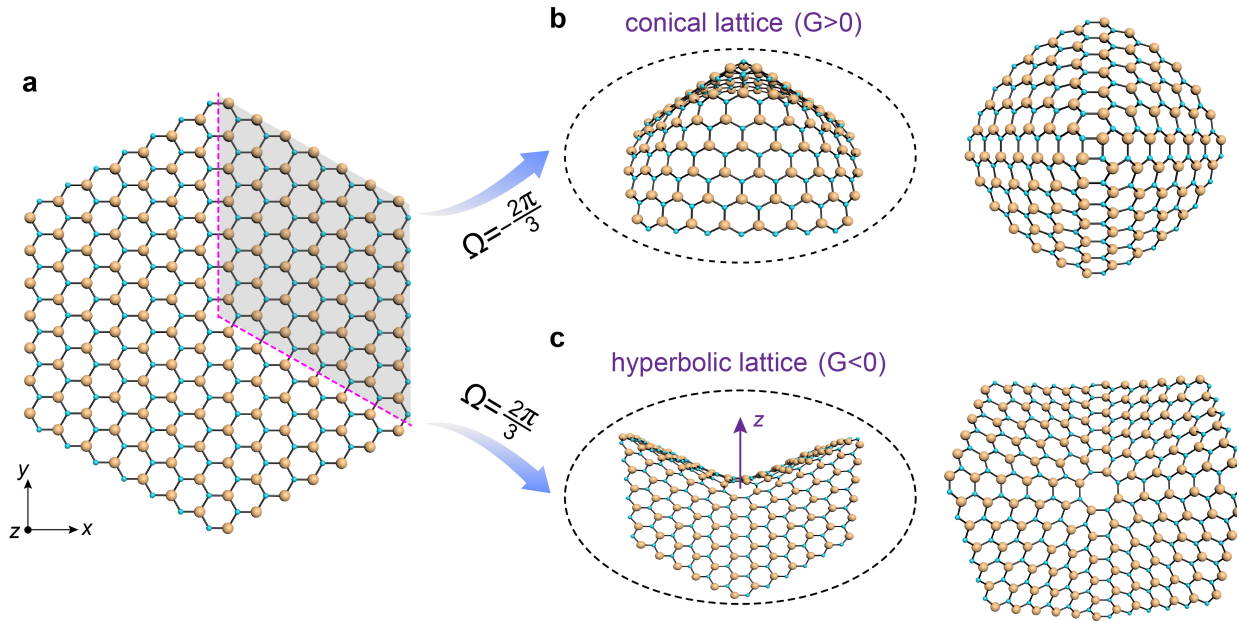

**Figure 1 | Non-Euclidean disclination structures. a,** A 2D honeycomb lattice consists of the A-type (yellow) and B-type (blue) sublattice sites. The shadow area is a $2\pi/3$ sector. **b-c,** Schematics of the deformed conical (hyperbolic) lattice constructed by removing (adding) a $2\pi/3$ sector from (into) a perfect lattice, i.e., the Frank angle of the disclination is $\Omega = -\frac{2\pi}{3}$ ($\frac{2\pi}{3}$). The left panels give the 3D view of the two disclinations, while the right panels give the top view of the structures. G represents the Gaussian curvature of the curved surfaces.

In Fig. 2b, we describe the acoustic system with a tight-binding model which is established by considering the σ-type nearest-neighboring couplings for the $p$-orbitals, while the π-type couplings are neglected. The nearest-neighboring couplings are obtained by examining the geometry of the $p$-orbitals and the hopping directions. Specifically, the hopping amplitude between two orbitals are determined by their projections into the hopping direction, since only the longitudinal (σ-type) couplings are significant (see Supplementary Note 1). It is thus important to introduce the three nearest-neighboring hopping directions from the A site (yellow) to its nearest-neighboring B sites (blue):

$$\vec{e}_{1,2} = \left( \pm \frac{\sqrt{3}}{2}, \frac{1}{2} \right), \vec{e}_3 = (0, -1). \tag{1}$$

By constructing and diagonalizing the tight-binding Hamiltonian within the six-component $p$-orbital spinor basis $[p_{A_x}(k), p_{A_y}(k), p_{A_z}(k), p_{B_x}(k), p_{B_y}(k), p_{B_z}(k)]^{\mathrm{T}}$, the band structure of the tight-binding model can be obtained, as shown in Figs. 2c and 2d (see details in Supplementary Note 1). When the on-site potentials, $\varepsilon_{A_j}$ and $\varepsilon_{B_j}$ ($j=x, y, z$), for the $p$-orbitals are zero, the band structure is given in Fig. 2c. At the $K$ point, there is a Dirac point which overlaps with two flat bands. Here, the Dirac cones as well as the upper and lower flat bands (purple curves) are due to the $p_x$- and $p_y$-orbitals, while the degenerate flat bands (yellow curves) overlapping with the Dirac points are due to the $p_z$-orbitals. When the onsite potentials are nonzero, e.g., $\varepsilon_{Ax} = \varepsilon_{Ay}$ = -1.8, $\varepsilon_{Az}$ = -2.1 and $\varepsilon_{Bx} = \varepsilon_{By}$ = 2, $\varepsilon_{Bz}$ = 2.8, both the Dirac cones and the degenerate flat bands are gapped (see Fig. 2d). The band gap of the Dirac cones is from -1.8 to 2, which manifests that they originate from the $p_x$- and $p_y$-orbitals as the gap is determined by $\varepsilon_{Ax,y}$ and $\varepsilon_{Bx,y}$. On the other hand, the flat bands are shifted to -2.1 and 2.8, which indicates that they originate from the $p_z$-orbitals and determined by on-site potentials $\varepsilon_{Az}$ and $\varepsilon_{Bz}$. The gapped band structure in Fig. 2d is due to the broken inversion symmetry and leads to the non-equivalent valley degrees of freedom in momentum space [43, 44]. Such a gapped band structure describes a valley Hall topological insulator emerging from $p$-orbital physics.

**Acoustic band structures and topology**

We construct the ball-and-stick sonic crystal of honeycomb lattice to realize the above tight-binding model and the non-Euclidean geometries due to the disclination structures. The acoustic band structures for the cases with and without the onsite potentials are shown in Figs. 2e and 2f. These band structures are obtained from acoustic simulation and are comparable with the tight-binding band structures in Figs. 2c and 2d, respectively. The unit-cell structure used in the acoustic simulation is shown in the inset of Fig. 2f. In this work, we set the distance between the nearest-neighboring spheres as $a$ = 31mm. The radii of the spherical cavities at the A and B sites are denoted as $r_A$ and $r_B$, respectively, while the diameter of the connecting tubes is set as $d$ = 5.6 mm. All these geometrical parameters are for the air regions encapsulated by the resins in 3D printing samples. Here, the on-site potentials can be controlled by the radii $r_A$ and $r_B$ for the A and B sites, respectively.

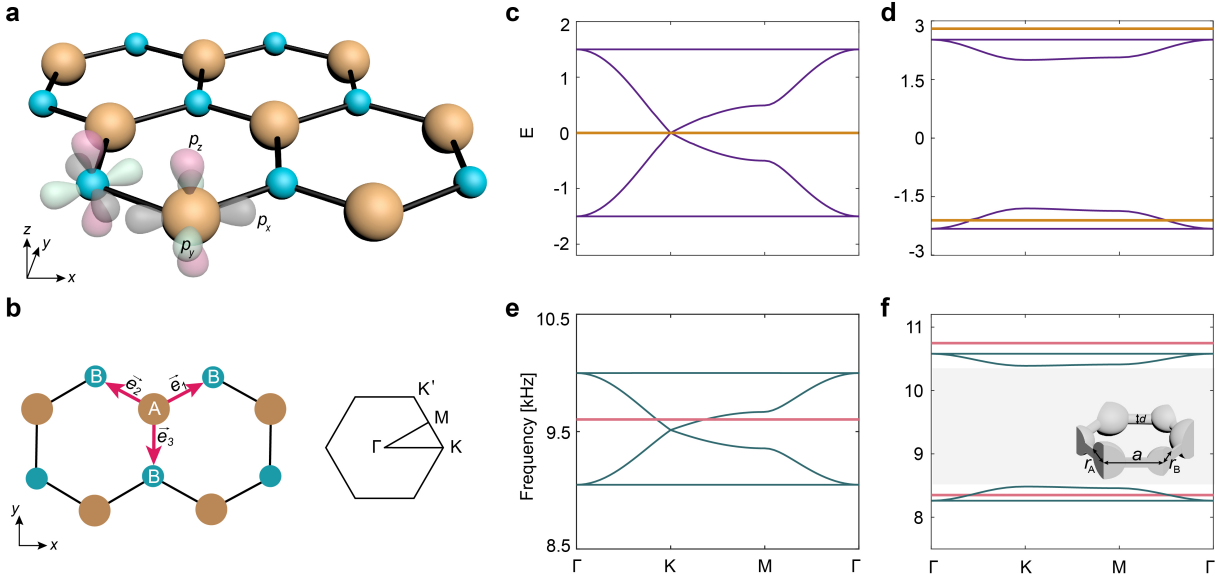

**Figure 2 | The _p_-orbital honeycomb lattice model in tight-binding and acoustic systems. a,** Illustration of the _p_-orbital honeycomb lattice model, including the $p_x$, $p_y$ and $p_z$ orbitals. **b,** Left: three hopping vectors $\vec{e}_1$, $\vec{e}_2$ and $\vec{e}_3$ from site A to its neighboring sites B. Right: the hexagon Brillouin zone. **c-d,** Band structures of the _p_-orbital tight-binding model along high symmetry lines for the coupling parameters $\varepsilon_{Ax,y,z}$ =0 and $\varepsilon_{Bx,y,z}$ =0 (**c**); while $\varepsilon_{Ax,y}$ =-1.8, $\varepsilon_{Az}$ =-2.1 and $\varepsilon_{Bx,y}$ = 2, $\varepsilon_{Bz}$ = 2.8 (**d**). **e-f,** Band structures of the ball-and-stick sonic crystal with its unit cell displayed in the inset of (**f**). The geometric parameters are $a$=31.0 mm, $d$ = 5.6 mm, and $r_A = r_B$ = 12 mm for the degenerate case (**e**); while $r_A$ = 13.5 mm, $r_B$ = 10.5 mm for the gapped case (**f**), with its topological gap denoted by the grey region.

For $r_A = r_B$ = 12 mm, the acoustic band structure gives the gapless Dirac cones at the _K_ and _K'_ points as well as the four flat bands (see Fig. 2e), resembling the band structure of the tight-binding model in Fig. 2c. Here, the difference is the two flat bands due to the $p_z$ orbitals do not overlap with the Dirac points, which reflects the difference in the on-site potentials for the $p_z$ and the $p_{x,y}$ orbitals. Such a difference is probably due to that the connecting tubes are in the _x-y_ plane which effectively modify the resonance frequencies of the $p_{x,y}$ orbitals. By breaking the inversion symmetry via setting $r_A$ = 13.5 mm and $r_B$ = 10.5 mm, the Dirac cones can be gapped, leading a valley Hall topological band gap ranging from 8.4 kHz to 10.43 kHz.

Meanwhile, the two degenerate flat bands due to the $p_z$ orbitals are split due to the on-site potentials, which are consistent with the tight-binding results in Fig. 2d.

We use the symmetry indicators to characterize the topology of the acoustic band gap in Fig. 2f. Here, the sonic crystal has $C_3$ rotation symmetry and the time-reversal symmetry. Following Ref. [45], the band topology can be characterized via the symmetry representations of the occupied bands at the high symmetry points (HSPs). By denoting the HSP as $\Pi$, the $C_3$ eigenvalues at $\Pi$ can be expressed as $\Pi_p=e^{2\pi i(p-1)/3}$ with $p = 1, 2, 3$, which can be identified from the phase profiles of the acoustic pressure field in the simulations. The topological index is given by $\chi = ([K_1], [K_2])$, where the integer topological invariants $[K_p] = \#K_p - \#\Gamma_p$. Here, $\#K_p$ ($\#\Gamma_p$) is the number of bands below the bandgap at the $K$ ($\Gamma$) point with the $C_3$ eigenvalues $K_p$ ($\Gamma_p$). We find that the topological index for the acoustic band gap in Fig. 2f is $\chi = (-2, 1)$ (see Supplementary Note 3 for details). Direct consequences of this topological band gap are the topological modes bound to the disclination cores in non-Euclidean geometry (denoted as the disclination modes), the valley edge states in finite systems with Euclidean geometry (see Supplementary Note 2), and a fractional mode charge at the disclination core. The fractional disclination charge $Q_{dis}$ is completely determined by the geometry of the disclination and the topological index as [25]

$$Q_{dis} = -\frac{\Omega}{2\pi}[K_1] \bmod 1, \qquad (2)$$

where $\Omega$ is the Frank angle. For the non-Euclidean geometry with $G > 0$ ($G < 0$), the Frank angle is $\Omega = -2\pi/3$ ($2\pi/3$). Here, the fractional disclination charge $Q_{dis}$ for acoustic systems should be interpreted as the integration of the local density-of-states of phonons for all the bulk bands below the band gap [25]. The above equation gives that the fractional disclination charge $Q_{dis} = 1/3$ ($2/3$) for the non-Euclidean geometry with $G > 0$ ($G < 0$). These fractional disclination charges are distinct from the fractional disclination charges revealed previously in photonic and electric circuit systems [28, 29].

**$p$-orbital disclination states in conical lattice**

To study the disclination states, we first construct an acoustic disclination in the conical lattice with 72 unit-cells (see inset of Fig. 3a). The simulated acoustic eigen-spectrum of this structure is shown in Fig. 3a (see Supplementary Note 1 for the calculation results of the tight-binding counterpart). There are 144 sites in this supercell and hence 432 eigenstates from the *p*-orbital bands. Four disclination states (denoted as group I and group II in Fig. 3a) emerge in the topological band gap. Note that the conical lattice has an emergent $C_2$ rotation symmetry around the disclination center. This emergent rotation symmetry, which is incompatible with the $C_3$ rotation symmetry of the original honeycomb lattice model in flat geometry, is a unique feature of the non-Euclidean geometry of our system. Due to the emergent $C_2$ rotation symmetry of the conical lattice, there are a symmetric (denoted as *S*) state and an anti-symmetric (denoted as *A*) state in both groups I and II. The acoustic wavefunctions of these calculated eigenstates are shown in Figs. 3b-3e from a top view. These wavefunctions manifest strong topological localization at the disclination center and rich *p*-orbital features.

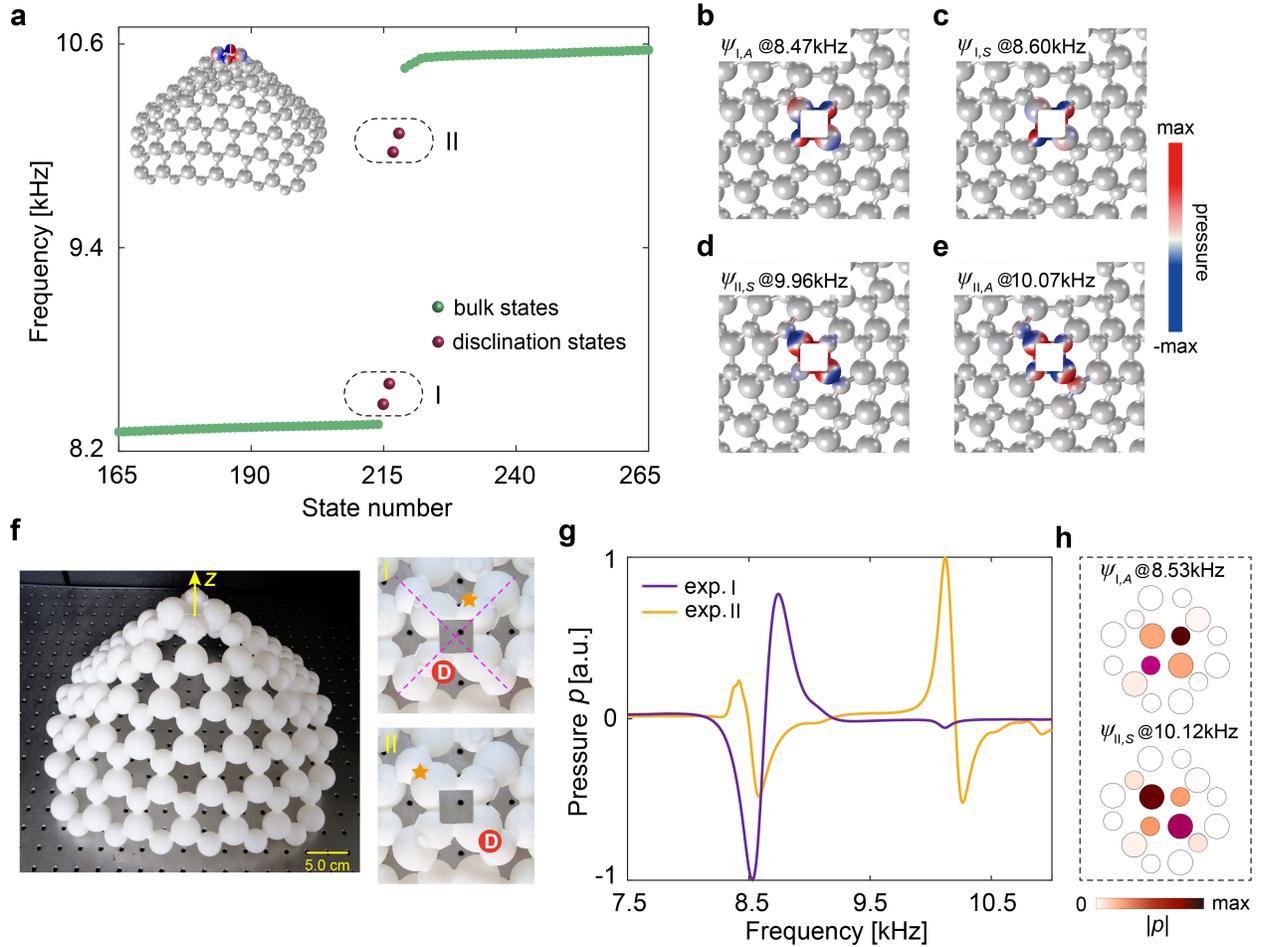

**Figure 3 | Observation of *p*-orbital disclination states in an acoustic conical lattice. a**, The calculated acoustic eigenfrequencies around the topological band gap in the conical lattice with $r_A$ = 13.5 mm, $r_B$

= 10.5 mm, and $d$ = 5.6 mm. The topological disclination states (groups I and II) are labeled by the red dots. The inset shows the acoustic pressure distribution of a disclination state in the conical lattice, which focuses mainly on the cavities at the system center. **b-e**, Top views of the acoustic pressure distributions of the disclination states in the $x$-$y$ plane. Note that our design of the unit-cell leaves partial cuts at the system center. **f**, Photo of the fabricated acoustic structure for the conical lattice. In the right panels, the pump-probe configurations I and II are shown where the orange stars (red circles) denote the positions of the acoustic source (detector). **g**, Measured acoustic pressure $p$ for the pump-probe configurations I and II at different frequencies. **h**, Measured acoustic pressure profiles $|p|$ (the absolute value) around the disclination core for two disclination states, $\Psi_{I,A}$ and $\Psi_{II,S}$, at their resonant excitation frequencies.

For example, the two disclination states in group I have wavefunctions that are highly concentrated on two small cavities at the disclination core, and the direction of the $p$-orbitals is perpendicular to the bond directions, i.e., features of π-like bonds (Figs. 3b-3c, corresponding to the eigen-frequencies of 8.47 kHz and 8.60 kHz, respectively). In contrast, for the disclination states in group II, the wavefunctions have different features: They are localized more at the large cavities at the disclination core, instead of the small cavities. Moreover, in these large cavities, the direction of the $p$-orbitals is along the bond directions (i.e., features of σ-like bonds), as shown in Figs. 3d-3e (corresponding to the eigen-frequencies of 9.96 kHz and 10.07 kHz, respectively). Overall, these in-gap disclination states manifest clear features that they originate from the $p_x$ and $p_y$ orbitals.

To experimentally confirm the above $p$-orbital disclination states, the acoustic structure is fabricated using commercial 3D printing technology based on photosensitive resins. A photo of the fabricated structure is shown in Fig. 3f. On each sphere around the disclination core, there are two or three small holes where the excitation speaker and the detection microphone can be inserted. Their positions are marked by the orange stars and the red circles, respectively. During the measurements, when some holes are free, rubber plugs are used to cover them to prevent sound wave leakage. Two pump–probe configurations, as shown in insets I and II of Fig. 3f, are used to detect the disclination states in groups I and II based on their wavefunction characteristics.

The detected acoustic signals for the pump–probe setups I and II are presented in Fig. 3g. The results show several peaks and dips that agree well with the calculated eigenfrequencies of the disclination states in Fig. 3a. In the pump–probe setup I, a dip and a peak at 8.53 kHz and 8.74 kHz, respectively, are observed which correspond to the two disclination states in group I, $\Psi_{I,A}$ and $\Psi_{I,S}$. For the pump–probe setup II, two resonances showing a Fano profile emerge at 10.12 kHz and 10.26 kHz, corresponding to the two disclination states in group II, $\Psi_{II,S}$ and $\Psi_{II,A}$. In setup II, the two disclination states in group I are also observed as a dip and a peak which is because the source and the detector in setup II also have overlap with the acoustic wavefunctions of the two disclination states in group I. We note that the pump-probe responses switch from positive to negative at the two resonances in both setup I and setup II which are essentially because of the switch between the inversion symmetry of the wavefunctions. For instance, in setup I, the lowest resonance is a dip because the source and detector overlap with the eigenstate with opposite wave amplitude. In contrast, the second resonance is a peak because the source and detector overlap with the eigenstate with the same amplitude. There are small blueshifts for the observed resonances when compared with the eigenstates frequencies from calculation which may be due to unavoidable fabrication errors in the sample.

To further verify the disclination states, we measure their acoustic wavefunctions when they are resonantly excited. Fig. 3h presents the absolute value of the detected acoustic pressure field at the cavities around the disclination core for the two states $\Psi_{I,A}$ and $\Psi_{I,S}$. Although the spatial resolution within each cavity is not achieved, these measured acoustic pressure profiles confirm the calculated acoustic wavefunctions (Here, we schematically depict the cavities as hollow circles). In particular, they demonstrate clearly that the group I eigenstates concentrate more on the small cavities, whereas the group II eigenstates concentrate more on the large cavities. Therefore, these observations confirm that the observed in-gap resonances are associated with the disclination states. More data are shown in the Supplementary Notes 4 and 6.

**$p$-orbital disclination states in hyperbolic lattice**

For the hyperbolic lattice, we consider a supercell with 100 unit-cells and the simulated acoustic eigen-spectrum is illustrated in Fig. 4a. There are in total 600 $p$-orbital eigenstates where some of them around the topological band gap are shown in the figure. We note that the hyperbolic lattice has an emergent $S_4$ improper rotation symmetry around the z axis at the disclination center. Again, this emergent rotation symmetry, which is incompatible with the $C_3$ symmetry of the honeycomb lattice model in flat geometry, is a feature of the non-Euclidean geometry here. From the eigen-spectrum, we find four states in the band gap. They form two non-degenerate symmetric modes ($\Psi_{S1}$ and $\Psi_{S2}$ at 8.98 kHz and 9.28 kHz, respectively) and two doubly degenerate anti-symmetric modes ($\Psi_{A1}$ and $\Psi_{A2}$ at 9.12 kHz), as required by the $S_4$ symmetry. The acoustic wavefunctions for the four in-gap disclination states are shown in Figs. 4b-4e. We note that the wavefunctions of these four states all concentrate more on the small cavities around the disclination core, showing features of $\pi$-like bonds of different symmetries.

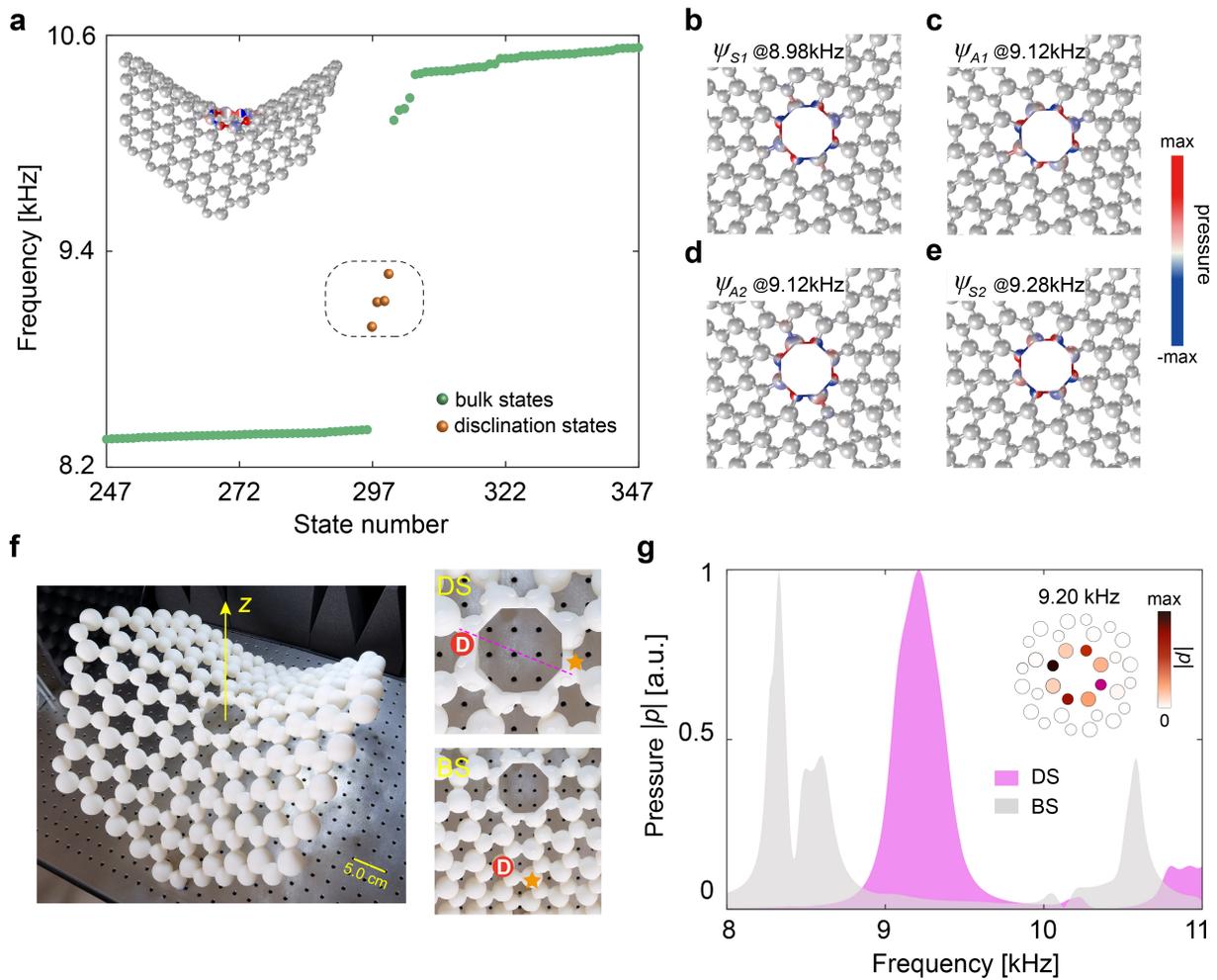

**Figure 4 | Observation of *p*-orbital disclination states in an acoustic hyperbolic lattice. a,** The simulated acoustic eigenfrequencies around the topological band gap in the hyperbolic lattice with $r_A$

= 13.5 mm, $r_B$ = 10.5 mm, and $d$ = 5.6 mm. The topological disclination states are labeled by the orange dots. The inset shows the acoustic pressure distribution of a disclination state in the hyperbolic lattice, which is mainly localized around the disclination core. **b-e**, Top views of the acoustic pressure distributions of the four in-gap disclination states. **f**, Photo of the fabricated acoustic structure for the hyperbolic lattice. In the right panels, the pump-probe configurations for the disclination and bulk states are shown (labeled as DS and BS, respectively) where the orange stars (red circles) denote the positions of the acoustic source (detector). **g**, Measured acoustic pressure $p$ for the disclination and bulk pump-probe configurations at different frequencies. **h**, Measured acoustic pressure profiles $|p|$ (the absolute value) for the disclination and bulk pump-probe configurations (labeled as DS and BS, respectively). The inset gives the measured acoustic pressure profiles for the cavities around the disclination core at the excitation frequency of 9.20 kHz (the peak of the violet spectrum).

To detect these four disclination states in experiments, we fabricate the hyperbolic lattice using the 3D printing technology. Fig. 4f shows the fabricated acoustic sample and the two pump-probe configurations that are used to detect the bulk and disclination states. In the bulk pump-probe configuration (denoted as BS in Figs. 4f-4g), both the source and the detector are away from the disclination core. In the disclination pump-probe configuration (denoted as DS in Figs. 4f-4g), the source and the detector are at the disclination core: The source at one side, while the detector at the opposite side. Therefore, only the disclination states can mediate the acoustic signal propagations from the source to the detector.

The measured acoustic responses for the two configurations are shown in Fig. 4g. The bulk response has a suppressed region from 8.8 kHz to 10.4 kHz which is due to the topological band gap in Fig. 2. In comparison, the disclination response has a broadened resonance around 9.2 kHz which is comparable with the frequencies of the four disclination states in Fig. 4a. Here, again, the small blueshifts of the resonance frequencies are possibly due to the fabrication errors in the sample. The measurement of the acoustic wavefunctions at their resonant excitation conditions are performed while the results are partly shown in the inset of Fig. 4g (other results are shown in Supplementary Note 5). From the figure, which shows the measured acoustic wavefunction at the excitation of 9.20 kHz, we find that the wavefunction is concentrated more on the small cavities around the disclination core which is consistent with the main feature of the disclination states in Figs. 4b-4e. The above measurements thus confirm the emergence of the disclination states in the topological band gap in the hyperbolic lattice. Finally, we note that other localized states out of the topological band gap are not discussed here for their trivial characteristics (see more details in Supplementary Notes 5 and 6).

**Discussions and outlook**

We theoretically propose and experimentally observe $p$-orbital disclination states in non-Euclidean geometries: The conical and hyperbolic lattices with disclinations of different Frank angles where the emergent $C_2$ and $S_4$ symmetries play important roles. Moreover, we unveil the rich orbital configurations of the topological disclination states. Our work reveals the intriguing interplay between the geometry and the band topology and extends the study of topological states to non-Euclidean curved surfaces. Lattice defects with finite Gaussian curvatures widely exist in nanostructures such as graphene and boron-nitride nano-cones [37]. The study here may inspire the exploration of the nontrivial electronic and phononic states in these nanostructures with non-Euclidean geometries. In addition, non-Euclidean geometries frequently appear in the virtual spaces in transformation optics and hence our study can be inspiring for the study of transformation optics [46]. More related topics for future research may include the Hofstadter butterflies in non-Euclidean geometries [47] and the extensions of our findings here to photonic, phononic and electrical circuit systems at various scales where non-Euclidean geometries are feasible in experiments.

# References

[1] Mermin, N. D. The topological theory of defects in ordered media. *Rev. Mod. Phys.* **51**, 591–648 (1979).

[2] Kleman, M & Friedel, J. Disclinations, dislocations, and continuous defects: A reappraisal.

 *Rev. Mod. Phys.* **80**, 61 (2008).

[3] Kosterlitz, J. M. Nobel lecture: topological defects and phase transitions. *Rev. Mod. Phys*. **89**, 040501 (2017).

[4] Iijima, S., Ichihashi, T. & Ando, Y. Pentagons, heptagons and negative curvature in graphite microtubule growth. *Nature*, **356**, 776-778 (1992).

[5] Lammert, P. E. & Crespi, V. H. Topological phases in graphitic cones. *Phys. Rev. Lett.* **85**, 5190 (2000).

[6] Karousis, N., Suarez-Martinez, I., Ewels. C. P. & Tagmatarchis, N. Structure, Properties, functionalization, and applications of carbon nanohorns. *Chem. Rev.*, **116**, 4850 (2016).

[7] Pachos, J. K., Stone, M. & Temme, K. Graphene with geometrically induced vorticity. *Phys. Rev. Lett.* **100**, 156806 (2008).

[8] Cortijo, A. & Vomediano, M. A. H. Effects of topological defects and local curvature on the electronic properties of planar graphene. *Nucl. Phys. B* **807**, 659-660 (2009).

[9] Vozmediano, M. A. H., Katsnelson, M. I. & Guinea, F. Gauge fields in graphene. *Phys. Rep.* **496**, 109-148 (2010).

[10] Yazyev, O. V. & Louie, S. G. Topological defects in graphene: Dislocations and grain boundaries. *Phys. Rev. B* **81**, 195420 (2010).

[11] Ran, Y., Zhang, Y. & Vishwanath, A. One-dimensional topologically protected modes in topological insulators with lattice dislocations. *Nat. Phys.* **5**, 298-303 (2009).

[12] Teo, J. C. Y. & Kane, C. L. Topological defects and gapless modes in insulators and superconductors. *Phys. Rev. B* **82**, 115120 (2010).

[13] Juricic, V., Mesaros, A., Slager, R. J. & Zaanen, J. Universal probes of two-dimensional topological insulators: dislocation and $\pi$-flux. *Phys. Rev. Lett.* **108**, 106403 (2012).

[14] De Juan, F., Ruegg, A. & Lee, D. H. Bulk-defect correspondence in particle-hole symmetric insulators and semimetals. *Phys. Rev. B* **89**, 161117 (2014).

[15] Slager, R.-J., Mesaros, A., Juričić, V. & Zaanen, J. Interplay between electronic topology and crystal symmetry: dislocation-line modes in topological band insulators. *Phys. Rev. B* **90**, 241403 (2014).

[16] Paulose, J., Chen, B. G. & Vitelli, V. Topological modes bound to dislocations in mechanical metamaterials. *Nat. Phys.* **11**, 153-156 (2015).

[17] Sumiyoshi, H. & Fujimoto, S. Torsional Chiral Magnetic Effect in a Weyl Semimetal with a Topological Defect. *Phys. Rev. Lett.* **116**, 166601 (2016).

[18] Liu, J. P. & Balents, L. Anomalous Hall effect and topological defects in antiferromagnetic Weyl semimetals: Mn3Sn/Ge. *Phys. Rev. Lett.* **119**, 087202 (2017).

[19] Chernodub, M. N. & Zubkov, M. A. Chiral anomaly in Dirac semimetals due to dislocations. *Phys. Rev. B* **95**, 115410 (2017).

[20] van Miert, G. & Ortix, C. Dislocation charges reveal two-dimensional topological crystalline invariants. *Phys. Rev. B* **97**, 201111 (2018).

[21] Li, F.-F. et al. Topological light-trapping on a dislocation. *Nat. Comm.* **9**, 2462 (2018).

[22] Slager, R.-J. The translational side of topological band insulators. *J. Phys. Chem. Solids* **128**, 24-38 (2019).

[23] Nayak, A. K. et al. Resolving the topological classification of bismuth with topological defects. *Sci. Adv.* **5**, eaax6996 (2019).

[24] Queiroz, R., Fulga, I. C., Avraham, N., Beidenkopf, H. & Cano, J. Partial lattice defects in higher-order topological insulators. *Phys. Rev. Lett.* **123**, 266802 (2019).

[25] Li, T. H., Zhu, P. H., Benalcazar, W. A. & Hughes, T. L. Fractional disclination charge in two-dimensional $C_n$-symmetric topological crystalline insulators. *Phys. Rev. B* **101**, 115115 (2020).

[26] Garrido, R. S., Muñoz, E. & Juričić, V. Dislocation defect as a bulk probe of monopole charge of multi-Weyl semimetals. *Phys. Rev. Research* **2**, 012043 (2020).

[27] Roy, B & Juričić, V. Dislocation as a bulk probe of higher-order topological insulators. *Phys. Rev. Research* **3**, 033107 (2021).

[28] Liu, Y. et al. Bulk-disclination correspondence in topological crystalline insulators. *Nature* **589**, 381-385 (2021).

[29] Peterson, C. W., Li, T. H., Jiang, W. T., Hughes, T. L. & Bahl, G. Trapped fractional charges at bulk defects in topological insulators. *Nature* **589**, 376-380 (2021).

[30] Wang, Q. et al. Vortex states in an acoustic Weyl crystal with a topological lattice defect. *Nat. Comm.* **12**, 3654 (2021).

[31] Xie, B. Y., You, O. & Zhang, S. Topological disclination pump. Preprint at https://arXiv.org/abs/2104.02852 (2021).

[32] Deng, Y. et al. Observation of degenerate zero-energy topological states at disclinations in an acoustic lattice. Preprint at https://arXiv.org/abs/2112.05182 (2021).

[33] Wang, Q., Xue, H. R., Zhang, B. L. & Chong, Y. D. Observation of protected photonic edge states induced by real-space topological lattice defects. *Phys. Rev. Lett.* **124**, 243602 (2020).

[34] Lin, Z.-K., Wu, Y., Jiang, B., Liu, Y., Wu, S., Li, F. & Jiang, J.-H. Topological Wannier cycles induced by sub-unit-cell artificial gauge flux in a sonic crystal. *Nat. Mater.* (2022).

[35] Ye, L. et al. Topological dislocation modes in three-dimensional acoustic topological insulators. Preprint at https://arXiv.org/abs/2104.04172 (2021).

[36] Yamada, S. S. et al. Bound states at partial dislocation defects in multipole higher-order topological insulators. Preprint at https://arXiv.org/abs/2105.01050 (2021).

[37] Gupta, S. & Saxena, A. A topological twist on materials science. *MRS Bull.* **39**, 265–279 (2014).

[38] Kamada, T. & Kawai, S. An algorithm for drawing general undirected graphs. *Inf. Process. Lett.* **31**, 7-15 (1989).

[39] Irvine, W. T. M., Vitelli, V. & Chaikin, P. M. Pleats in crystals on curved surfaces. *Nature* **468**, 947-951 (2010).

[40] Matsumoto, E. A. et al. Wrinkles and splay conspire to give positive disclinations negative

curvature. *Proc. Natl. Acad. Sci.* (USA) **112**, 12639 (2015).

[41] Wu, C. J. & Das Sarma, S. $p_{x,y}$-orbital counterpart of graphene: Cold atoms in the honeycomb optical lattice. *Phys. Rev. B* **77**, 235107 (2008).

[42] Lu, X. C., Chen, Y. & Chen, H. Y. Orbital corner states on breathing kagome lattices. *Phys. Rev. B* **101**, 195143 (2020).

[43] Rycerz, A., Tworzydlo, J. & Beenakker, C. W. J. Valley filter and valley valve in graphene. *Nat. Phys.* **3**, 172-175 (2007).

[44] Xiao, D., Yao, W. & Niu, Q. Valley-contrasting physics in graphene: Magnetic moment and topological transport, *Phys. Rev. Lett.* **99**, 236809 (2007).

[45] Beanalcazar, W. A., Li, T. & Hughes, T. L. Quantization of fractional corner charge in $C_n$-symmetric higher-order topological crystalline insulators. *Phys. Rev. B* **99**, 245151 (2019).

[46] Zhao, P., Cai, G. & Chen, H. Exact transformation optics by using electrostatics. *Sci. Bull.* **67**, 246-255 (2022).

[47] Stegmaier, A., Upreti, L. K., Thomale, R. & Boettcher, I. Universality of Hofstadter butterflies on hyperbolic lattices. Preprint at https://arXiv.org/abs/2111.05779 (2021).

## Methods

### Simulations

The acoustic band structures and eigenfrequencies are performed by the finite-element based software COMSOL Multiphysics with the pressure acoustic module. For the acoustic waves in 3D printed acoustic crystal structures using photosensitive resin, all air-solid interfaces are modeled as sound hard boundaries. Sound waves propagating in air with a mass density of 1.25 kg/m$^3$ at the speed of 343 m/s are set in simulations. The bulk band structure of the 3D acoustic crystal (Fig. 2e, f in the main text) are calculated using a single unit cell with periodic boundary conditions in *x-y* planes. For the acoustic eigenstates in Fig. 3 and Fig. 4, two supercells with outer sound hard boundaries (the inset in Fig. 3a and 4a) are established in simulations.

## Experiments

Our experimental samples were manufactured by stereolithographic 3D printing technology using photosensitive resins (modulus 2460 MPa, density 1.1 g/cm3) with a fabrication tolerance of roughly 0.1 mm. The design geometry in Fig. 2 is used for the air cavities and the connecting tubes which are encapsulated by the resin layer with a thickness of 3 mm. To measure the acoustic signals inside the cavity, holes with a diameter of 8 mm was left on each excitation and detection cavity. When measuring the acoustic fields in experiment, we kept the holes in all other cavities (except for the source and detection position) closed by plastic plugs. An acoustic source (tiny speaker) was placed into the cavity, which is connected to a waveform generator (RIGOL DG4062) and powered by an amplifier to launch broadband acoustic signals with the frequency sweeping from 7 kHz to 11 kHz at a step of 10 Hz. A probe microphone with diameter of ~7mm, which can easily be put into the cavities through holes are used to measure the acoustic signals. Then, the data are collected by a DAQ card (NI USB-6361), meanwhile the amplitude and phase distributions of the acoustic pressure fields can be extracted by Fourier Transform. To ensure data correctness, the signal at each position is averaged out of five repeated measurements to reduce the noise interference. As shown in Fig. 3, 4 in the main text, in experiment, we finally use different acoustic pump-probe measurements to detect the signals of localized disclination states in our samples.

## Acknowledgements

Y.C., Y.Y. and H.C. are supported by the National Natural Science Foundation of China (Grant Nos.12104169, 92050102 and 11874311), the Fundamental Research Funds for the Central Universities (Grant No. 20720200074). Y.L., Z.-K. L. and J.-H.J. acknowledge support from the National Natural Science Foundation of China (Grant No. 12074281). Y. C. acknowledges support from Huaqiao University (605-50Y21003).

## Author contributions

Y.C., J.-H.J., and H.C. conceived the idea. Y.C., Y.L., and Z.-K. L. performed the theoretical analysis and the numerical simulations. Y.Y. and Z.-H.Z. designed and performed the experiments.

Y.C., J.-H.J, and H.C. supervised the project. All the authors contributed to the discussions of the results. J.-H.J and Y.C. wrote the manuscript and the Supplementary Information.

## Competing Interests

The authors declare no competing interests.

## Data availability

All data are available in the manuscript and the Supplementary Information. Additional information is available from the corresponding authors through reasonable request.

## Code availability

We use the commercial software COMSOL MULTIPHYSICS to perform the acoustic wave simulations and eigenstates calculations. Reasonable request to computation details can be addressed to the corresponding authors.